\documentclass[amsmath,amssymb,twocolumn]{revtex4}
\usepackage[parfill]{parskip}    
\usepackage{graphicx}
\usepackage{amssymb}
\usepackage{epstopdf}
\usepackage{color}
\usepackage{graphicx}
\usepackage{pdfpages}

\begin{document}

\title{A self-sustained traversable scale-dependent wormhole}
\author{Ernesto Contreras {${}^{a}$} \footnote{On leave 
from Universidad Central de Venezuela}
\footnote{ej.contreras@uniandes.edu.co} and 
Pedro Bargue\~no{${}^{a}$}\footnote{p.bargueno@uniandes.edu.co}}
\address{${}^a$Departamento de F\'{\i}sica,
Universidad de los Andes, Apartado A\'ereo {\it 4976}, Bogot\'a, Distrito Capital, Colombia}

\begin{abstract}
A self--sustained traversable wormhole is obtained as a vacuum solution of a scale--dependent gravitational
theory. Comparison with other approaches towards wormhole self--sustainability are presented, with emphasis
on the running of the gravitational coupling and on a possible effective description of gravity near the Planck scale.
\end{abstract}

\maketitle


\section{Introduction}\label{intro}

Wormholes are bridges between different universes or different parts of
the same universe. They were first recognized by Flamm, who found that
the Schwarzschild solution of can be thought as representing a wormhole \cite{Flamm1916}. Later, Einstein and Rosen 
\cite{Einstein1935} developed a model for an elementary particle consisting in a throat connecting two sheets.
After the ``mass without mass"--like elaborations of Wheeler and Misner in their Geometrodynamics \cite{Wheeler}, the field
experimented a renaissance due to the work of Morris and Thorne \cite{Morris1987,Morris1988}. In the last 20 years, the book by
Visser \cite{Visserbook} has been considered as an appropriate and authoritative reference in the field. Very recently, Lobo
\cite{Lobo2017} has also authoritatively updated on the state--of--the--art on wormhole physics. Therefore, after having a look at 
this timeline, one can conclude that we are living a new renaissance on wormholes and related physics. 

Let us focus our interest in traversable wormholes. As it is well known, the problem with them is that they violate the classical 
energy conditions, serving primarily as useful probes of the foundations of General Relativity. The kind of matter which allows 
traversable wormholes is called {\it exotic}. As a consequence, quantum effects must be considered in order to solve the 
{\it traversability problem}. 
As the full theory of quantum gravity is still lacking,
many different works have been devoted to get some insight into the underlying physics (for an incomplete list check 
\cite{Deser:1976eh,Rovelli:1997yv,Bombelli:1987aa,Ashtekar:2004vs,Sakharov:1967pk,Jacobson:1995ab,Verlinde:2010hp,Reuter:1996cp,
Litim:2003vp,Horava:2009uw,Charmousis:2009tc,Ashtekar:1981sf,Penrose:1986ca,Connes:1996gi,Nicolini:2008aj,Gambini:2004vz} and for a 
review see \cite{Kiefer:2005uk}). Despite the fact that in those works the authors discuss different aspects of 
quantum gravity, most of them have the common feature that the resulting effective gravitational action acquires a scale--dependence. 
This behaviour is observed through the couplings of the effective action: they change from fixed values to scale--dependent 
quantities, i.e. $\{G_0, \Lambda_0\} \mapsto \{G_k, \Lambda_k\}$, where $G_0$ is Newton's coupling and $\Lambda_0$ is the 
cosmological coupling. Indeed, there 
is some evidence which supports that this scaling behaviour is consistent with Weinberg's Asymptotic Safety program 
\cite{Weinberg:1979,Wetterich:1992yh,Dou:1997fg,Souma:1999at,Reuter:2001ag,Fischer:2006fz,Percacci:2007sz,Litim:2008tt}. 
This effective action which appears when running couplings are assumed has been studied in three--dimensional
space--times in the context of black hole physics \cite{Rincon:2017goj,Rincon:2017ypd,koch2016,Rincon2018} 
as well as in four dimensions
\cite{Koch:2014joa,Koch:2014joab,Contreras2018}. In these works, the corresponding scale--dependent couplings take into account a quantum effect in the sense
that this approach admits corrections to the classical black hole backgrounds. 

From the point of view of Semiclassical General Relativity, self--consistent solutions to the semiclassical Einstein's equations
corresponding to a Lorentzian wormhole coupled with a quantum scalar field  have been considered by Hochberg {\it et al.} 
\cite{Hochberg1997} and by Khusnutdinov and Sushkov \cite{Khusnutdinov2002}. 
Regarding self--sustainability, Garattini fixed the attention on wormholes which are totally supported by their own quantum 
fluctuations
(see the original works \cite{Garattini2005,Garattini2007} and also \cite{Lobo2017} and references therein for a recent account 
of these kind of wormholes). By studying the one--loop contribution of the gravitons to the total energy, which is quite 
similar to computing the Casimir energy on a fixed background, he found a self--consistent source for a traversable wormhole
\cite{Garattini2005}. An important feature of this self--sustainability lies in the fact that a renormalized energy--dependent 
Newton's gravitational constants appears as a consequence of considering effective Einstein's equations coming from the
fluctuations of the Einstein tensor. Therefore, in this sense, an effective action description of self--sustaina\break ble wormholes 
should be possible. This description, within a scale--dependent gravitational setting, is the purpose of the present work.

The work is organized as follows: In Sect. \ref{einstein_w} we give a brief review on traversable wormholes in Einstein's gravity.
Section \ref{scale_theory} summarizes the
gravitational scale--dependent setting which is employed in Sect. \ref{scale_wormhole} to obtain a self--sustained 
wormhole with the Schwarzschild spatial part of the metric. Finally, discussion and concluding remarks are given in Sects. \ref{discussion} and \ref{remarks},
respectively.

\section{Traversables wormholes in Einstein's gravity}\label{einstein_w}

Let us consider a Morris--Thorne wormhole \cite{Morris1987}, which is one of the simplest traversable wormholes. 
It can be described by a 
static and spherically symmetric line
element given by 
\begin{eqnarray}
\label{wmetric}
ds^{2}=-e^{2 A(r)}dt^{2}+\frac{dr^{2}}{1-\frac{B(r)}{r}}+r^{2}d\Omega^{2}. 
\end{eqnarray}
For a static observer, the only nonzero components of the stress-energy tensor are
\begin{eqnarray}\label{stress}
T^{t}_{t}&=&\rho(r)\nonumber\\
T^{r}_{r}&=&-\tau(r)\nonumber\\
T^{\theta}_{\theta}&=&T^{\phi}_{\phi}=P(r),
\end{eqnarray}
where $\rho(r)$ is the total density of mass--energy, $\tau(r)$ is the tension per unit of area in the radial direction and
$P(r)$ is the pressure in lateral directions.

With the above parametrization of the line element and the choice (\ref{stress}) for the matter content, the Einstein's field
equations lead to
\begin{eqnarray}
\rho&=&\frac{B'}{8\pi r^2}\\
\tau&=&\frac{B/r-2(r-B)A'}{8\pi r^{2}}\\
P&=&\frac{r}{2}\bigg((\rho-\tau)A'-\tau'\bigg)-\tau.
\end{eqnarray}

The former equations suggest that, for a suitable choice of the functions
$A(r)$ and $B(r)$, we can obtain the matter contain for our wormhole.

However, the functions $A(r)$ and $B(r)$ are not arbitrary but they must fulfill some constraints in order to obtain a
traversable wormhole.
For example, if there is not cutoff in the stress--energy we must demand that \cite{Morris1987,Morris1988}
\begin{eqnarray}
\frac{B}{r}\rightarrow 0\\
A(r)\rightarrow0,
\end{eqnarray}
as $r\rightarrow\infty$. Furthermore, the requirement that a traversa\break ble wormhole does not possess any horizon corresponds
to demand $A(r)$ to be finite everywhere. 

As an example, we will briefly comment on two types of traversable wormholes of ultrastatic 
\footnote{A spacetime is ultrastatic if can be written in some coordinate system as $g=-dt^2+g_{ij}dx^{i}dx^{j}$} type.
 
First, the prototype of traversable wormhole, 
which is the Ellis--Bronnikov one \cite{Ellis1973,Bronnikov1973,Ellis1979}. This
wormhole has a shape function given by $B(r)=r_{0}^2/r$ (with $r_{0}$ constant) and $A(r)=0$. 
Note that, although these wormholes were thought to be 
unstable \cite{Shinkai2002,Gonzalez2009a,Gonzalez2009b,Torii2013}, 
rotation might possibly stabilize them \cite{Matos2006}. Even more, Bronnikov {\it et al.} have shown
\cite{Bronnikov2013} that a perfect fluid with negative density and a source-free radial electric or magnetic field 
(for a certain class of 
fluid equations of state) allows linear stability for the Ellis--Bronnikov solution under both spherically symmetric 
perturbations and axial perturbations of arbitrary multipolarity (see also Bronnikov's study on Chapter 7 of \cite{Lobo2017}).
Very recently, and in analogy with black holes \cite{Heusler1998}, uniqueness theorems for
the Ellis--Bronnikov wormhole supported by a phantom scalar field has been proven both in four \cite{Yazadjiev2017} 
and in higher--dimensional cases \cite{Rogatko2018}.
Concerning the study of gravitational lensing of wormholes, due to their astrophysical importance, 
the deflection of light for Ellis--Bronnikov wormholes was initially computed in \cite{Chetouani1984}. Other authors have 
extended the study of these
kind of signatures both in non--rotating \cite{Abe2010,Abe2011,Taka2013,Zhou2016,Tsukamoto2016,Tsukamoto2017,Tsukamoto2017b} 
and rotating Ellis--Bronnikov wormholes \cite{Jusufi2018}.

Second, let us consider
a wormhole with $A(r)=0$ and $B(r)=r_{0}$. As pointed out by Morris and Thorne \cite{Morris1987}, the
parameter $\xi=\frac{\tau}{\rho}-1$ quantifies the amount of exotic material needed to sustain the wormhole. In this particular
case, although the exotic material decays rapidly with radius, $\xi$ is positive and huge. In this sense,
the authors of Ref. \cite{Morris1987} point out that this situation, which implies the use of exotic material throughout all the
wormhole, is extremely unpleasing. 
Given this unpleasant feature of these wormholes when interpreted whithin General Relativity, 
in this work we will show that they are an exact {\it vacuum} solution of a particular scale--dependent gravity. Therefore,
no exotic matter but a modified gravitational theory is implemented in order to obtain a self--sustained wormhole solution of 
this type.

\section{Scale--dependent gravity}\label{scale_theory}

As commented in the introduction, one possible way of introducing an effective gravitational theory beyond General Relativity is
by promoting both the Newton and the comological constants to scale--dependent quantities. In the following, the scale--setting 
presented will follow closely the spirit and concept of Ref. \cite{Koch:2014joa}.

The scale--dependent Einstein--Hilbert effective action reads
\begin{eqnarray}\label{action}
\Gamma[g_{\mu\nu},k]=\int \mathrm{d}^{4}x\sqrt{-g}\bigg[\frac{1}{16\pi G_{k}} (R-2\Lambda_{k}) 
+\mathcal{L}^{\text{M}}_k\bigg],
\end{eqnarray}
where $G_{k}$ and $\Lambda_{k}$ stand for the scale--dependent
 gravitational and cosmological coupling, respectively, and $\mathcal{L}_{k}^{\text{M}}$ is the Lagrangian density for the 
matter content.\\
After performing variations with respect to the metric field $g_{\mu\nu}$, we obtain 
the modified Einstein's field equations 
\begin{eqnarray}\label{einstein}
G_{\mu\nu}+g_{\mu\nu}\Lambda_{k}=8\pi G_{k}T^{eff}_{\mu\nu},
\end{eqnarray}
where
$T^{eff}_{\mu\nu}$ is the effective energy--momentum tensor, defined as
\begin{eqnarray}\label{eff}
T^{eff}_{\mu\nu}:=(T^{\text{M}}_{k})_{\mu\nu} - \frac{1}{8\pi G_{k}}\Delta t_{\mu\nu}.
\end{eqnarray}
In Eq. (\ref{eff}), $(T^{\text{M}}_{k})_{\mu\nu}$ is the matter energy--momentum tensor and $\Delta t_{\mu\nu}$ is given by
\begin{eqnarray}\label{nme}
\Delta t_{\mu\nu}=G_{k}\left(g_{\mu\nu}\square -\nabla_{\mu}\nabla_{\nu}\right)G_{k}^{-1}.
\end{eqnarray}
As discussed previously in Ref. \cite{Rincon:2017goj}, the renormalization scale $k$ is not constant anymore. Therefore, 
the stress energy tensor is likely not conserved. This kind of problem 
has been considered in the context of renormalization group improvement of black holes in asymptotic safety 
scenarios (see, for instance \cite{Bonanno:2000ep,Bonanno:2006eu,Reuter:2010xb,Koch:2014cqa} and references therein). 

One can circumvent this problem by applying the variational scale--setting procedure described in Ref. \cite{Koch:2014joa},
where the Eqs. (\ref{einstein}) are complemented by an equation obtained performing variations with respect to the scale--field, 
$k(x)$:
\begin{align}\label{scale}
\frac{\mathrm{d}}{\mathrm{d} k} \Gamma[g_{\mu \nu}, k] =0.
\end{align}

However, if the precise beta functions of the problem are not known, 
 Eqs. (\ref{action}) and (\ref{scale}) do not provide enough information in order to find both $g_{\mu\nu}(x)$ and $k(x)$.
One can solve this problem by considering that the couplings $\{G_k$, $\Lambda_k\}$ depend explicitly on space--time coordinates,
a dependence which is inherited from the space-time dependence of $k(x)$
\cite{Koch:2014joa,Koch:2014joab,koch2016,Rincon:2017goj,Rincon:2017ypd}. Promoting Newton's coupling to a space--dependent
field, $G(x)$, and finding wormhole solutions for this modified theory, is the purpose of the following section. Note that
this scale--dependent gravity corresponds to an effective Brans--Dicke theory but without a kinetic term. In this sense, 
$G(x)$ does not have dynamics.

\section{Self--sustained scale--dependent solution}\label{scale_wormhole}

The modified vacuum Einstein's equations without cosmological term are given by $S_{\mu\nu}=G_{\mu\nu}+\Delta t_{\mu\nu}=0$, where
\begin{eqnarray}
S^{t}_{t}&=&-2 G^{2}(r) B'(r)+4 r (r-B(r)) (G'(r))^2+ \nonumber \\
&+&G(r) \bigg(\left(3 B(r)+r \left(-4+B'(r)\right)\right) G'(r) \nonumber \\
&+&2 r (-r+B(r)) G''(r)\bigg)\label{Stt} \\
S^{r}_{r}&=& G(r) \left(-B(r)+2 r (r-B(r)) A'(r)\right)\nonumber \\
&+&r (-r+B(r)) \left(2+r A'(r)\right) G'(r) \label{Srr}\\
S^{\theta}_{\theta}&=&S^{\phi}_{\phi}=
B(r)\bigg(-4 r^2 G'(r)^2-G^{2}(r) \bigg(-1+r \bigg(A'(r) \nonumber \\
&+&2 r (A'(r))^2 
+2 r A''(r)\bigg)\bigg) 
+r G(r) \bigg(\left(1+2 r A'(r)\right) \times \nonumber \\
&\times&G'(r) +2 r G''(r)\bigg)\bigg)
r \bigg(4 r^2 (G'(r))^2+G^{2}(r) \bigg(\bigg(1+ \nonumber \\
&&r A'(r)\bigg) \left(2 r A'(r)-B'(r)\right)+
2 r^2 A''(r)\bigg)+r G(r) \times \nonumber \\
&\times &\bigg(\bigg(-2 -2 r A'(r)
+B'(r)\bigg) G'(r)-2 r G''(r)\bigg)\bigg) \label{Sthetatheta}
\end{eqnarray}

Note that the equations are highly coupled. However, the following protocole can be implemented in order to look for
some solutions. First,
solve for $G'(r)$ from Eq. (\ref{Srr}). Second, substitute $G'(r)$ in Eq. (\ref{Stt}) and then solve for $G''(r)$. With these
algebraic identities, Eq. (\ref{Sthetatheta}) results in
\begin{eqnarray}
\label{eqlong}
&&A'(r) \bigg(-B(r) \left(6+r A'(r) \left(1+r A'(r)\right)\right) \nonumber \\
&+&r \left(r A'(r) \left(2+r A'(r)-B'(r)\right)
-2 \left(-8+B'(r)\right)\right)\bigg) \nonumber \\
&+&8 B'(r)+2 r (r-B(r)) \left(4+r A'(r)\right) A''(r)=0.
\end{eqnarray}
Surprisingly, Eq. (\ref{eqlong}) does not contain $G(r)$. Even more, one possible solution is given, by inspection, by
\begin{eqnarray}
A(r)&=&A_{0} \\
B(r)&=&B_{0}.
\end{eqnarray} 

Note that, given this choice for the metric, all the equations (where $G(r)$ is the only unknown) can be consistently solved 
leading to
\begin{equation}
\label{Gdr}
G(r)=\frac{G_{0}}{\sqrt{1-\frac{B_{0}}{r}}},
\end{equation}
where $G_{0}$ is the classical Newton's constant. It is worth noticing that, as the redshift is constant ($A_{0}$), 
the radial tidal acceleration felt by an observer trying to traverse the wormhole is zero. On the contrary, the 
transversal tidal acceleration essentially depends on the velocity with which the observer traverses the wormhole 
\cite{Morris1987,Visserbook,Lobo2017}.

\section{Discussion}
\label{discussion}
At this point, a number of comments are in order. First, note that the spatial part of the obtained wormhole is similar
to that of a Schwarzschild wormhole (the Schwarzschild redshift is somehow incorporated in the effective $G(r)$). 
Second, 
in the context of scale--dependent gravity, the wormhole throat, $B_{0}$, can be interpreted as the so--called running parameter,
which controls the strength of the scale--dependence \cite{Koch:2014joa,Koch:2014joab,koch2016,Rincon:2017goj,Rincon:2017ypd}.
In other words, when the running parameter is turned off, $B_{0}\rightarrow 0$, the
classical solution is recovered, and $G(r)\rightarrow G_{0}$. Even more, this limit corresponds to
Minkowski spacetime, as can be easily checked. In this sense, the solution here presented can be considered to be
self--sustained by a
scale--dependent gravitational theory where the effective Newton's constant is given by Eq. (\ref{Gdr}). 
Third, as in general
the scale--dependent effects are assumed to be weak \cite{Koch:2014joab}, it is reasonable to treat 
the running parameter, 
which we recall is encoded in 
$B_{0}$, as small with respect to the other scales entering the problem. Therefore,
the effects of the running of $G(r)$ \footnote {$G(r)$ can be considered to be an effective beta function in the
spirit of \cite{Koch:2014joab}.} are expected to be noticeable only near the throat. Specifically, as $[B_{0}]=L$ and
$[G_{0}]=[L]^{1/2}$ when $c=\hbar=1$, we get that $B_{0} < \sqrt{G_{0}}=l_{p}$. Then, provided scale--dependent
gravity can be considered as an effective model for quantum gravity in some sense, the (trans)--planckian
bound obtained for $B_{0}$ is consistent and, even more, it is in agreement with
\cite{Khusnutdinov2002,Garattini2005,Garattini2007}.

Within this interpretation,
no violation of energy conditions appears, since we are dealing with a vacuum spacetime, but a modified gravity emerges.
In fact, as pointed out in Ref. \cite{Lobo2017}: ``in the context of modified theories of gravity, it is shown that the 
higher--order curvature terms, interpreted as a gravitational fluid, can effectively sustain wormhole geometries, while the 
matter threading the wormhole can be imposed to satisfy the energy conditions". In our case, the matter
content which Ref. \cite{Lobo2017} refers to is the vacuum and the new gravity is not given by higher--order curvature terms
but by the scale--dependence. Therefore, scale--dependent gravity provides a possible realization
of the previous claim.
 
Concerning the self--sustainability of the wormhole note that, one one hand, 
in the approach of Refs. \cite{Garattini2005,Garattini2007}, the
effective Einstein's equations are given by
\begin{equation}
G_{\mu\nu}=-\langle \Delta G_{\mu\nu}(\bar g_{\mu\nu}, h_{\mu\nu})\rangle^{\mathrm{ren}},
\end{equation}
where $-\langle \Delta G_{\mu\nu}(\bar g_{\mu\nu}, h_{\mu\nu})\rangle^{\mathrm{ren}}$ is an effective energy--mom\break entum
tensor which appears as a consequence of a one--loop 
renormalization procedure over a fixed wormhole background given by $g_{\mu\nu}$ ($\bar g_{\mu\nu}=g_{\mu\nu}+h_{\mu\nu}$). Moreover,
given the fact that an arbitrary mass scale, $\mu$, emerges unavoidably in any regularization scheme, a scale--dependent running
gravitational coupling appears. The specific running obtained in Ref. \cite{Garattini2005} reads
\begin{equation}
\label{new}
G(\mu)=\frac{G_{0}}{1+K G_{0}\ln(\mu/\mu_{0})},
\end{equation}
where $K$ is a constant related to the background geometry and $\mu_{0}$ is the normalization point.
On the
other hand, within
our approach, the effective Einstein's equations are given by
\begin{equation}
G_{\mu\nu}=-\Delta t_{\mu\nu},
\end{equation}
where the effective energy--momentum tensor appears when the scale--dependence can not be avoided anymore. Therefore, one can
conclude that the scale--depen\break dence of the gravitational coupling provides an effective mechanism for the inclussion of 
quantum effects in the context of wormholes. In this sense, the obtained solution can be also taken to be self--sustainable, 
but this time due to the effect of the running of the Newton's gravitational coupling.


\section{Concluding remarks}\label{remarks}

In this work we have constructed the first wormhole solution in the context of scale-dependent gravity. 
Interestingly, the
obtained geometry is a vacuum solution of the modified Einstein's equations and, therefore, no violations of the energy 
conditions appear. The width of the wormhole's throat has been shown to correspond to the running parameter, which measures
deviations from General Relativity. Even more, this parameter controls the running of the Newton's coupling, which appears
to be redshifted instead being constant as in the usual case. We have noted that this wormhole is self--sustained in the sense
that the obtained effective gravity is the only responsible of its sustainability. In this sense, the model here
presented can be thought as an effective description of previously considered self--sustained wormholes, 
which is confirmed by their (trans)--planckian size.
Therefore, following \cite{Garattini2005,Garattini2007}, we conclude that the obtained traversability has to be regarded as 
in ``principle" rather than in ``practice". 
Finally, in order to propose some astrophysical signatures of the wormhole here presented, a study of its stability is mandatory.
We leave this and other topics for a future work.

\section*{ACKNOWLEDGEMENTS}
The author P. B. was supported by the Faculty of Science and Vicerrector\'{\i}a de Investigaciones of Universidad de los Andes, 
Bogot\'a, Colombia. P. B. dedicates this work to Ana\'{\i}s Dorta--Urra and to Luc\'{\i}a and In\'es Bargue\~no--Dorta.

\end{document}